# Porous radiation dielectric detectors.

Lorikyan M.P.
Yerevan Physics Institute

Abstract

Multiwire and microstrip porous detectors have been developed and investigated for DC operation. The multiwire porous detector consists of anode wires, an Al cathode and a gap between them filled with porous CsI. The microstrip porous detector consists of an insulating plate covered with metallic strips, micromesh anode and a gap between them filled with porous CsI. For some time after being manufactured, these detectors' performances are non-stable and they have poor spatial resolution. However, after being kept in vacuum for a certain time, they spontaneously acquire stability and spatial resolution better than 100 μm and have detection efficiencies of 100% and 70% for heavily ionizing α-particles and 5.9 keV X-rays respectively. The MWPDD performs stably at an intensity of heavily ionizing α –particles of 711cm$^{-2}$ s$^{-1}$.

# 1. Introduction

The development of many areas of science, engineering, biology and medicine are tightly bound (interlinked) to methods of registration of particles and X-rays. Therefore, the search for new methods of registration of radiation has been a priority direction in physics. One of the methods developed for ionizing radiation detection during the last 30 years is the method of porous dielectric detectors (PDD). In this detector, electrons drift and multiply (EDM) in porous dielectrics (PD) under the influence of an external electric field. Success in this direction came from groups at the Yerevan Physics Institute where that phenomenon was observed and porous detectors were developed and systematically studied [1-5]. The multiplication of electrons takes place because of secondary electron emission from the walls of pores allowing porous detectors to have good time and spatial resolutions. For porous detectors, active materials such as KCl, CsI, KBr, $Na_3AlF_6$, MgO and others having a density of a few percent of the bulk density were used. The EDM in PD with an external electric field has been demonstrated as controllable by the electric field and provides a very high yield $\sigma$ of secondary electrons per incident heavily or minimum ionizing particles. These detectors are very thin having an extremely small amount of matter in the path of the particle (~ 200μg/cm$^2$)[1-9]. Anomalous secondary electron emission (ASEE-Malter effect [10], which occurs in the field of positive charges accumulated on the surface of a porous emitter, also provides a very high secondary electron emission factor $\sigma$ for low energy and intensity electron beams [11-17]. Moreover, according to estimations of E.J. Sternglass et al., for 1-10 MeV α-particles and β-particles the ASEE $\sigma$ can reach $10^3$ and 4 respectively and the formation time should not exceed $0,5 \cdot 10^{-10}$ s [18]. In 1965 Garwin et al., [19,20] have shown that the ASEE factor for porous emitters in the case of a high energy electron beam has $\sigma \approx 5$ and for ≈ 2 MeV single β-particles $\sigma \approx 1$. Such small values of $\sigma$ for single weakly ionizing particles are caused by the method of creation of the electric field in the porous emitter. This method consists of the following: before irradiating with the particles to be detected, the emitter is bombarded by an intense low-energy (~ 10 keV) electron beam. The electron beam traversing the porous layer knocks secondary electrons out of the surface, with the remaining positive ions creating an electric field in the porous dielectric. The main drawback of this method is that simultaneously with the accumulation of the positive charges on the layer's surface, charges accumulate also inside this layer (on the surfaces of pores and inside the walls). These bulk charges capture electrons and holes produced by the radiation being detected and suppress the intensity of the EDM and SEE. They increase the fluctuations of SEE gain and introduce different kinds of instabilities, since density of those charges is directly related to the primary particles intensity. In the main, because of these disadvantages and owing to such an inconvenient method for the creation of an electric field in the porous emitter, ASEE is not used for single particle detection.

Having assumed, that all these problems do not arise when an external electric field is used, Lorikyan et al., [1,2] in the early 1970's tested porous secondary emitters with an applied external electric field. Using this method, the authors observed secondary electron emission (CSEE) [1-4] highly controllable by an external electric field. Besides, the use of an external field makes it possible to observe a SEE from thick layers by applying higher electric fields and have led to the development of a series of new radiation detectors. The emitter used in above mentioned experiments consisted of the thin metallic foil covered with porous KCl with relative density $\rho_{rel}$ = 2% of the bulk density and a fine metallic control grid which is located on the surface of KCl layer or is placed at some distance from this surface. An external electric field was applied between the cathode and control grid. Investigations have shown that for I ≈ $3.2 \cdot 10^{-10}$ A of 50 MeV incident electron beam traversing the emitter, the secondary electron yield $\sigma$ sharply increases with electric field and reaches $\sigma$ = 32. When the grid was located on the surface of a porous KCl film, secondary electrons appeared immediately after the beginning of bombardment, however, when the grid was not in contact with the film, the maximum value of $\sigma$ was reached in the course of time, i.e. in the last case inertial SEE is observed. However, the advantages of the external electric field were displayed in full when the porous emitter was shot through by a single minimum ionizing β-particle. Lorikyan et al. have shown that in this case, σ also grows sharply with the electric field E (see fig.1) but reaches a higher value (220) than in the previous case [3,4]. It is worth noting, that the mean free path λ of electrons in the porous layer for an electron beam is $\lambda \approx 1.0$ μm [1,2], but for single particles $\lambda \approx 10$ μm [3,4]. Such a large difference between the values of σ and λ for beams and single incident particles is due to the fact that beams produce a very large bulk charge in the porous layer while for single particles, the polarization charge is small. The relatively high average values of secondary electron energies of $\varepsilon \approx 60$ eV and $\varepsilon \approx 20$ eV for $E = 4\times10^4$ V/cm and $E = 2\times10^4$ V/cm [3,4] ensure the fast signal of porous detectors. One of the drawbacks of this method is that the statistical distributions of secondary electrons [3-5,21-23] for both weakly and heavily ionising particles are too wide and not suitable for energy measurements, although for 5.5 MeV α-particles 100% detection efficiency and $\sigma \approx 2\times10^3$ was obtained [5-9].

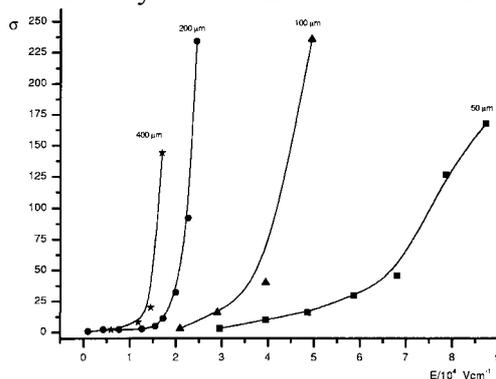

**Fig. 1** The dependences of σ on the electric field strength for various thicknesses of porous KCl layer for minimum ionizing β-particles traversing emitter [3,4].

Thus a high quality of operation of porous detector is ensured by: 1) applying the external electric field; 2) with the contacts to the anode and cathode electrodes on the surfaces of the porous layer; 3) relatively high secondary electron energies; 4) small sizes of pores (which limit the transverse size of the electron avalanche and determines the spatial resolution of the detector). These requirements were first met in multiwire and microstrip porous dielectric detectors (MWPDD) [24-34]. Using the obtained values of $\lambda \approx 10$ μm, $\varepsilon = 50$ eV, for 1.5-2.5 MeV β-particles the number of electrons collected on anode wires for MWPDD with thickness and density of porous KCl layer 300 μm and 2% respectively was estimated to be $M \sim 6\times10^4$ [34] the registration efficiencies of 5,5 MeV α-particles, minimum ionizing β-particles and 5.9 keV X-rays and their time and coordinate resolutions were shown to be very good [27,30]. The disadvantage of these detectors was that they operated stably only in the pulsed mode, i.e. when after a short time of operation (from several milliseconds to one second) the MWPDD is fed with a pulse of inverse polarity [28,29]. Detailed information on PDDs is available in ref [34].

A SEE of porous dielectric emitters was also studied in 1983 by Cheehab et al. [35] with $\sim 10^{-9}$A 25 MeV electrons beam and in 1989 by Chianelli et al., [36] using >1 MeV β-particles and 5,5 MeV α–particles. The emitters used in [38,39] had no metallic control grid on the surface of the porous layer and the anode electrode was not in contact with porous layer. In reference[36] when the voltage turned on, the σ for these particles was about 15 and 2000 respectively, and then it fell. In that work the detection efficiency of 0.95 for 540 MeV protons was achieved.

The high secondary emission properties of alkali-halides aroused intense interest and yields from various porous and solid CsI emitters were observed in gas atmospheres [37-52]. Of course, not only alkali-halides, but also any other efficient secondary electron emitters can be used as an active media of such detectors. A relevant example could be the so called micro sphere plates operating in the DC mode, in which pores are the spaces between the spheres [53-55].

In recent works [56-58] it was found that under certain conditions the performances of porous multiwire and microstrip detectors are stable even in the DC mode of operation and they have good coordinate resolution

This paper presents the results of a study in the DC mode of multiwire porous dielectric detectors (MWPDDs) and microstrip porous dielectric detectors (MSPDDs) filled with porous CsI. Most investigations were made for source intensities $I \leq 12$ cm$^{-2}$ sec$^{-1}$ and detectors had CsI with purity 99.99, while counting rate capabilities were studied for α –particles with intensity of $I = 711$ cm$^{-2}$ s$^{-1}$ and higher CsI layer purity (99.999). Studies with 5.5 MeV α –particles and 5,9 keV X- rays were performed. Numerous porous detector prototypes have been tested and the results of typical detectors are presented.

## 2. Principles of Operation of Porous Dielectric Detectors

Note that the phenomenon of EDM in porous dielectrics under the action of an external electric field have not been sufficiently well studied yet. The absence of a quantitative theory of the EDM in porous dielectrics under the influence of an external electric field seriously restricts the development of porous dielectric detectors and the possibility of interpretation of obtained results.

Qualitatively the process of EDM in porous dielectrics can be described as follows: in the pore walls of the material, primary particles create electrons in a conduction band and holes in the valence band. Electrons with energies higher than the electron affinity $\chi$ of the wall surface penetrate through the walls into the vacuum in the pores i.e. secondary electrons emission (SEE) arises. Secondary electrons are accelerated in the pores by the applied electric field up to energy of $\varepsilon = eE\lambda$, where $e$ is charge of electron, $\lambda$ is the electron's free path in porous medium and $E$ is the electric field strength. The accelerated electrons produce new secondary electrons and holes in the medium. This process repeats producing an avalanche multiplication of electrons (holes). A high value of $\sigma_t$ of dielectric emitters used in porous detectors is the result of a large band gap W of dielectrics (W = 4 - 8 eV) and relatively low electron affinity of these materials [59]. High values of W are necessary since in that case more electrons participate in EDM, which is due to the fact, that conducting band electrons with energy $\varepsilon < W$ (in slightly defective crystal lattice) do not interact with the valence-band electrons, hence their energy losses in dielectrics are low. As a result the mean energy of electrons in conducting band of dielectrics is high, the electron's penetration depth Z is large and the number of electrons reaching the pore surface with energy sufficient to overcome the surface barrier is high. As a consequence of the low electron affinity, more valence-band electrons exit into the pores and participate in the process of EDM. Table 1 gives the values of a W, Z, χ, and σ for some materials, (Z and σ are obtained for primary electron energies of 1-5 keV) [60]

Table 1.

| Material | BaO | Al$_2$O$_3$ | MgO | LiF | KB$_2$ | CsI | KI | NaCl | CsBr |
|---|---|---|---|---|---|---|---|---|---|
| σ | 1.05 | 1.3 | 3.8 | ~1 | 7.9 | 9.7 | 4.9 | ~8 | - |
| Z, Å | 230 | 240 | 320 | 20 | 50 | 900 | 250 | 460 | 1200 |
| W, eV | 3.7 | - | 8.7 | 12 | - | 6.3 | 6.2 | - | - |
| χ, eV | 1.3 | - | 1 | 1 | - | 0.1 | 1.1 | - | - |

From this table, we see that CsI has a high value of W and the lowest value of χ and therefore is suitable as an active medium for porous detectors. Using the value of the free path λ=10μm [4], one can estimate, that a porous CsI layer with a density of ≈1% that of solid CsI, the thickness of the walls is about 1000 Å, which is close to the value of the Z of CsI (see Table.1). This and the small value of the $\chi = 0.1$ eV are responsible for the high gain of the porous CsI layer in the external electric field.

However all aforesaid is not sufficient for complete description of the EDM processes in porous emitters. The defects in a crystalline structure and the impurities in a porous medium act as traps of electrons and holes. Trapped electrons and holes form bulk and surface charges, which reduce the EDM efficiency and spoil all characteristics of detectors. We can conclude that a dielectric porous detector with active medium having a noticeable density of traps has poor spatial and energy resolutions and unstable characteristics. That is why the purities of the material and of the procedure of detector manufacturing are very important [61]. The effect of polarization charges on counting characteristics of the porous detectors was observed already in our early studies [24-33].

### 3. Description and assembly of Porous Dielectric Detectors

#### 3.1. MWPDD

The MWPDD look like gaseous multiwire proportional chambers, where porous materials replace the gas. The design of the MWPDD is presented schematically in fig.2. The anode was made of the 25-micron diameter gold plated wires. The cathode was made of 60 μm rigid type of Al foil stretched on a fiberglass frame. Insulator spacer of 0.5 mm thickness set the gap between wires and cathode, which was filled with porous CsI. MWPDDs with different spacing between anode wires b were tested in these studies. α-particles and X-quanta enter into the porous medium from anode wires side and through the Al foil cathode respectively.

The MWPDD was assembled in this way: the gold plated wires were stretched between two screws with the help of a tension of 65g, and then they were fastened onto the fiberglass frame (anode frame) by epoxy. The cathode Al foil was stretched using two metallic round frames, and then cathode frame was glued to it.

The openings of anode and cathode frames determine the active area of detectors. For assembly of the MWPDD, a porous CsI with thickness L is sputtered on the cathode and then the anode wires were brought into contact with CsI layer and frames were fixed to each other.

Microscopic studies of the surface of the detector revealed that the immersion of the wires into the porous layer is very small, their action results mainly in the compression of the layer.

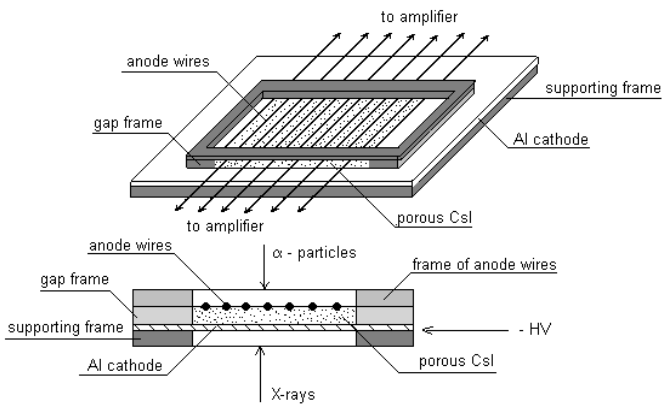 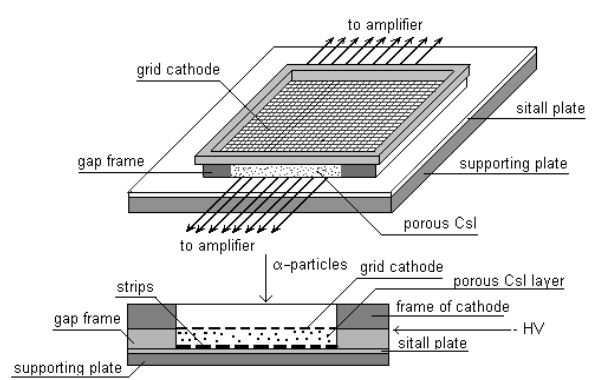

**Fig. 2** The schematic design of the MWPDD.    **Fig. 3** The schematic design of the MSPDD.

3.2. MSPDD

The design of the MSPDD is presented schematically in fig.3. In the MSPDD gold strips were lithographically deposited onto the glass-ceramic plate (readout plate)[1]. The width of strips was 20 μm, the distance between their centers was 100 μm. The resistance of each strip attained 50 Ω. The cathode was made of 0.7 transparency micromesh. An insulator spacer of 0.5 mm thickness sets the gap between strips and cathode and it was filled with porous CsI.

The MSPDD was assembled as follows: the cathode micromesh is stretched and glued on the cathode frame in the same manner as Al cathode foil of MWPDD. Primarily the readout plate was glued to the fiberglass substrate, after that porous CsI of thickness L was deposited on strips of the readout plate. The micromesh cathode was placed on the CsI layer and cathode frame was fixed to the readout plate substrate.

## 4. Description and production of porous CsI layer.

To provide a good technological cleanliness, all parts of the detectors were in advance cleansed chemically and were repeatedly washed in distilled water and dried at a room temperature.

The porous CsI layer is prepared by thermal deposition from a tantalum boat in the atmosphere of low argon pressure [38]. The boat was chemically cleansed beforehand. The relative density of the CsI layer just after deposition was $\rho_{rel} \approx 0.4\%$. This density is obtained for argon pressure p = 3 Torr, the distance from the boat to substrate H = 6.5 cm and deposition rate of 56 μm/min. The accuracy in determining of ρ was 12 %. The thickness of porous layer L was measured with micrometer in the ambient Ar immediately after the deposition. As the porous layer is very soft, the accuracy of measurement of L was poor ($\approx$ 50 μm). The non uniformity of the CsI layer thickness over the area just after deposition didn't exceed the measurement accuracy of 50 μm.

It was observed that the thickness of porous CsI layer kept in vacuum decreases with time. The results of measurements of the CsI layer thickness changes with time are given in fig.4. Since this decrease may result in the loss of the contact between the CsI and the cathode of the MSPDD (or anode wires of MWPDD), the CsI layer was prepared with a thickness L exceeding that of the detector gap by a factor of 1.5-2.

The compression of the CsI layer could take place also under the action of the external electric field. The effect of an electric field on the gap filled with porous CsI was studied by an optical

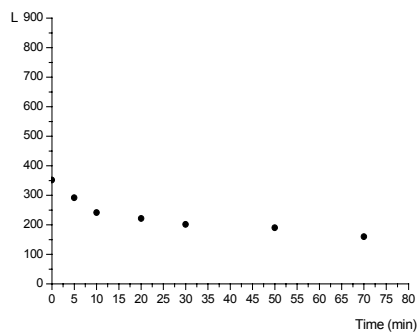

**Fig. 4** The results of measurements of CsI layer thickness variation with time after preparation.

---

[1] The readout plates were manufactured in the open joint-stock company, Moscow, RF.

microscope and for voltages up to 2 kV the changes of the gaps were less than 20μm.

**4.1 Detectors setting up in the device**

The emission properties of a porous layer deteriorates in time if it is in contact with air. To exclude such changes, the detectors which were assembled in the box with argon, were moved in argon to experimental setup and were quickly installed in the experimental vacuum chamber. The vacuum chamber, which was beforehand filled with argon, was then brought rapidly to a vacuum of $\approx 7 \times 10^{-3}$ Torr.

**5. Measurement procedure.**

The amplifier used for amplification of detector's pulses had a rise time of 1.5 ns, a 250-Ohm input impedance, and a conversion ratio of 30 mV/μA. After amplification the signals were discriminated with a threshold $V_{thr}$ and were registered by counters.

Both N(s), the numbers of particles detected by only one anode wire or strip and N(c), the number of particles detected simultaneously by two adjacent anode wires or strips were measured. The total number of detected particles is N(s+c) = N(s) + N(c). We denote the corresponding detection efficiencies by η(s), η(c) and η(s+c) respectively, η(s+c)= η(s) + η(c). This allows us to measure the detection efficiency of particles and simultaneously to evaluate the upper limit of the spatial resolution of detectors. To do that the anode wires of the MWPDD (or strips of the MSPDD) were divided into two groups (see fig. 5).

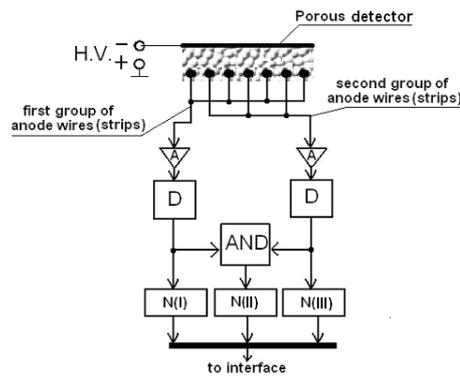

**Fig. 5** Schematic view of the partition of anode wires or strips into two groups and basic diagram for measuring the efficiency and coordinate resolution: A – amplifier; D – discriminator; AND – coincidence; N(I), N(II), N(III) – counters.

The first group included the even-numbered and the second group included the odd-numbered anode wires or strips. Anode wires (strips) in each group were connected with each other and numbers of signals from first group N(I) and second group N(II) were separately amplified and counted. The number of coincidence pulses from both groups of wires (strips) $N_c$ was also counted. Therefore, $N_c \equiv$ N(c), N(s) = N(I) + N(II) - 2N(c). It is clear that when in the porous dielectric detector the width of the electron shower is much smaller than b, the parameter Ss = N(c)/N(s+c) « 1 and detector spatial resolution is not worse than b. When the width of the electron shower in the porous dielectric detector is more than the anode wires (strip) separation b, then N(c) ≈ N(I) = N(II) and parameter Ss ≈ 1, (the detector's spatial resolution is worse than b).

The detectors were held at room temperature T and the vacuum was ≈ $7 \times 10^{-3}$ Torr. During time stability studies the detectors were usually switched off for 16 hours every day and always at weekends. In all cases measurements errors are only statistical and are not indicated in figures. When irradiation was stopped, the counting rates were ≈ 0.1- 0.2 $s^{-1}$.

## 6. Results of Investigation.

Observation of the shape of amplified pulses with 500 MHz oscilloscope have shown that the rise time, full duration and mean amplitude of pulses generated by α particles are ≈ 2.5 ns and ≈ 7.5 ns about 70 mV respectively.

The characteristics of the detectors used and the conditions of the experiment, such as the effective area of the detector $A$, pitch b, the distance d between a radiation source and detector, the thickness L of porous layers just after the deposition, the intensity I of a particles entering the porous layer, the temperature T at which the detector was kept after preparation and the purity of the CsI layer are given below in tables 2 -10.

### 6.1. Investigation of multiwire porous detectors
6.1.1. Registration of alpha particles.

**Experiment # 1**. *Study of MWPDD switched on in one hour after assembling.*

Parameters of the experiment                                            Table 2

| Detector | A, cm$^2$ | b, μm | d, cm | L, mm | I, cm$^{-2}$·s$^{-1}$ | T, $^0$C | Imurity of CsI |
|---|---|---|---|---|---|---|---|
| MWPDD | 2.2×2.2 | 250 | 12 | 0.8 | 8.8 | 18 | 10$^{-4}$ |

Investigations have shown, that for some time after being manufactured a porous CsI detector performance is not stable and it has poor spatial resolution. However, after being kept in vacuum for a certain time they perform stably and have good spatial resolution. The characteristics of MWPDD were first measured one hour after assembly. The MWPDD anode voltage U of the numbers of α - particles detected by first group anode wires ($N_\alpha$(I)), by second group wires ($N_\alpha$(II)) and simultaneously by both anode wires groups ($N_\alpha$(c)) are shown in fig.6.

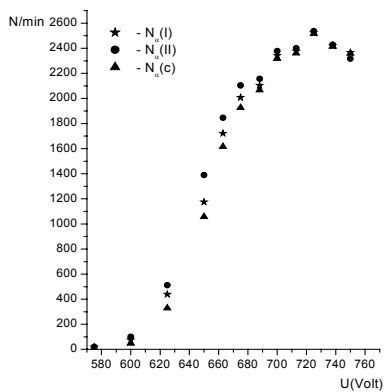 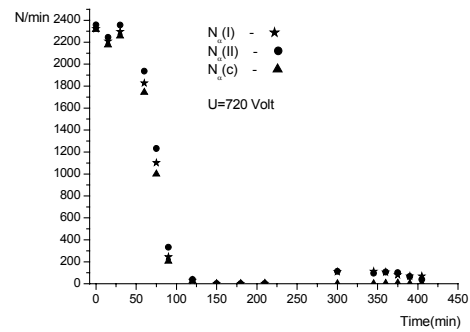

**Fig.6**          **Fig.7**

**Fig. 6** The variation with the MWPDD anode voltage U of the numbers of α – particles detected by first group anode wires $N_\alpha$(I), by second group wires $N_\alpha$(II) and by both wires groups simultaneously $N_\alpha$(c) obtained in one hour after assembly of the detector. The detector after assembly was kept in vacuum at temperature 18$^0$C.

**Fig. 7** The time stability of the numbers of α – particles detected by first group anode wires $N_\alpha$(I), by second group wires $N_\alpha$(II) and by both anode wires groups simultaneously $N_\alpha$(c) obtained just after finishing the measurements results of which are demonstrated in figure 6.

One can see that: 1) $N_\alpha$(I) and $N_\alpha$(II) sharply rise with voltage and then plateau. This is because the secondary emission factor σ is directly related to the electric field E; 2) within the experimental errors always $N_\alpha$(I) ≈ $N_\alpha$(II). This is normal because, as both groups of anode wires are identical and are located symmetrically with respect to the α-source; 3) $N_\alpha(c) \approx N_\alpha(I) \approx N_\alpha(II)$, all α-particles are detected simultaneously by both groups of anode wires, i.e. detector's spatial resolution at that time is worse than the anode-wire spacing b = 250 μm; 4) 4) the total detection efficiency of α-particles $\eta(s+c) = \eta(s) + \eta(c)$ is ≈ 0.96% ($N_\alpha(I) \approx N_\alpha(II) \approx N_\alpha(0) = 2562\,min^{-1}$). Investigations of the time stability of $N_\alpha(I)$, $N_\alpha(II)$ and $N_\alpha(c)$ made just after these measurements were finished show that the detector's performance at the time was unstable. The corresponding results are shown in fig.7.

Right after the completion of these measurements, $N_\alpha(I)$, $N_\alpha(II)$ and $N_\alpha(c)$ were again investigated and it was found that the counting characteristics of the detector improved dramatically but with a higher voltage required. Results of the variation with the anode voltage U of $N_\alpha(I)$, $N_\alpha(II)$, $N_\alpha(c)$ and the total number of registered α-particles $N_\alpha(s+c)$ are presented in fig.8.

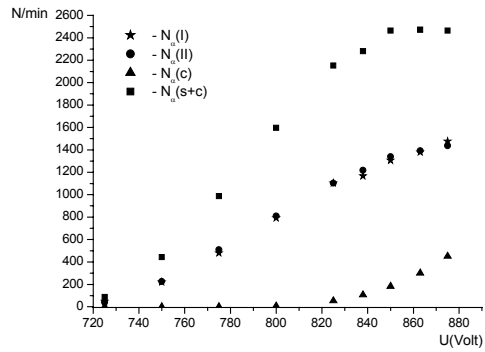

**Fig. 8** The variation with the MWPDD anode voltage U of the numbers of α – particles detected by first group anode wires $N_\alpha(I)$, by second group wires $N_\alpha(II)$, by both wires groups simultaneously $N_\alpha(c)$ and the total number of registered α – particles $N_\alpha = N_\alpha(s) + N_\alpha(c)$ obtained just after observation of results given in figure 7.

One can see that in this case also $N_\alpha(I) \approx N_\alpha(II)$, but $N_\alpha(c)$ is very small (Ss << 1), i.e. during 7 hours of operation the width of the electron shower in the porous dielectric detector became less that the pitch b = 250 μm, and spatial resolution of this detector became not worse than b = 250 μm. Comparing these results with results of fig.6 we see that in fig. 8: a) the dependences of $N_\alpha(I)$ and $N_\alpha(II)$ on U are shifted by 125 V to the right, b) both $N_\alpha(I)$ and $N_\alpha(II)$ are half those in fig.6. The first result can possibly be explained by the increase of the porous CsI layer density with time (see fig. 5), since that brings to a decrease of the sizes of pores and higher voltage U is needed to accelerate electrons up to the energy required for multiplication. The second result is due to the fact, that in the case of the fig.6 each group of anode wires registered all incident particles, while in the case of fig.8 each group of anode wires registered about half of these particles. From fig.8 one can also see, that in the range of U > 800V N(c) increases. The reason for that is clear because, as was shown experimentally [5], the SEE gain and the energy of shower electrons increases with an electric field strength (see fig.1), hence the share of shower electrons reaching adjacent wires simultaneously also grows. The time stability investigations made after completion of these measurements have shown that during the experiment the detector's performance was stable, detection efficiency was near 100 % and always $N_\alpha(c)$ was about 8 % of $N_\alpha(s+c)$.

The reason why one hour after the manufacture of a porous CsI PDD, the performance is non-stable and has poor spatial resolution is probably connected with the fact, that just after preparation by evaporation, the porous CsI layer has a high density of bulk and surface defects. These defects are charge-carrier trapping centers, which capture the electrons and holes produced by the incident radiation, thus forming high space and surface charges, which polarize the dielectric medium. The fields of these charges weaken and destroy the external electric field in PDD and as a result suppress the effectiveness of the EDM, widen the transversal sizes of the electron shower and deteriorate the spatial resolution of the detector.

The acquisition by MWPDD of spatial sensitivity and time stability probably can be explained within above described framework by a considerable decrease of densities of bulk and surface defects in porous medium.

Experiment # 2. *Study of MWPDD switched on in 14 hour after assembling*

Parameters of the experiment                                                                 Table 3

| Detector | A, cm$^2$ | b, μm | d, cm | L, mm | I, cm$^{-2}$·s$^{-1}$ | T, $^0$C | Impurity of CsI |
|---|---|---|---|---|---|---|---|
| MWPDD | 2.2×2.2 | 250 | 7.5 | 1.1 | 9.3 | 18 | 10$^{-4}$ |

This investigation shows that after being kept in vacuum for a certain time after assembly, a MWPDD performs stably and has good spatial resolution even if it was not switched on. For this problem the detector was switched on 14 hours after assembly. Figure 9 shows U dependences of total detection efficiency of α - particles η(s+c) = η(s) + η(c) and, the efficiency of simultaneous detection of α - particle by two group of anode wires $\eta_\alpha$(c).

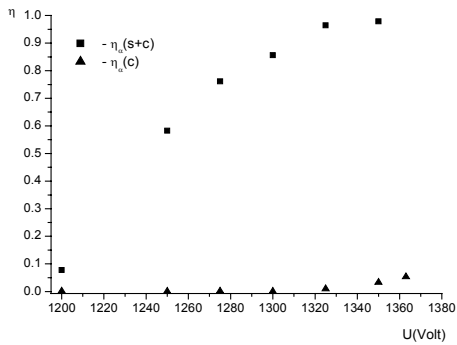 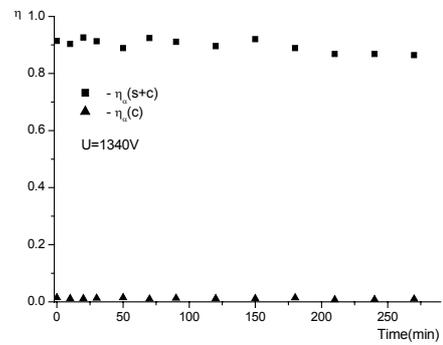

**Fig. 9**                                                                                              **Fig. 10**

**Fig. 9** U dependences of the total detection efficiency of α – particles $\eta_\alpha$(s + c) = $\eta_\alpha$(s) + $\eta_\alpha$(c) and the efficiency of simultaneous detection by the both groups of anode wires $\eta_\alpha$(c) for MWPDD which was switched on in 14 hour after assembly. The detector after assembly was kept in vacuum at temperature 18$^0$C.

**Fig. 10** The time stability of total detection efficiency of α - particles $\eta_\alpha$(s + c) = $\eta_\alpha$(s) + $\eta_\alpha$(c) and the efficiency of simultaneous detection by the both groups of anode wires $\eta_\alpha$(c) of MWPDD for 4.5 hours operation. These results were obtained just after observation of date illustrated in figure 9.

We see that always $\eta_\alpha$(c) « $\eta_\alpha$(s+c) and in a region of plateau Ss ≈ 0.1, i.e. after 14 hours of rest detector's spatial resolution became better than b = 250 μm. From fig.9 and fig.8 we can notice, that the curves η(s+c) = f(U) and $\eta_\alpha$(c) = f(U) in fig.9 are shifted to the right by approximately 500V compared with those in fig. 8. This effect is due to the thickness of the porous layer L of this MWPDD being larger than in the previous case. After assembly of the MWPDD, the porous layer was compressed by about 1.5 times more than in the case of L = 0.8 mm. Therefore the sizes of the pores are smaller than in the previous detector and a higher voltage U is needed to accelerate electrons to the energy required for multiplication. Afterwards, the detector operated stably and maintained good coordinate resolution. The results of the time stability study for 4.5 hours are shown in fig.10. These results indicate that the stabilization processes occur when after preparation MWPDD is kept in vacuum in an idle state.

**Experiment # 3.** *Study of MWPDD with a small pitch.*

Parameters of the experiment                                                                    Table 4

| Detector | A, cm$^2$ | b, μm | d, cm | L, mm | I, cm$^{-2}$·s$^{-1}$ | T,$^0$C | Impurity of CsI |
|---|---|---|---|---|---|---|---|
| MWPDD | 1×2.5 | 125 | 12 | 0.8 | 9.1 | 18 | 10$^{-4}$ |

To determine the transverse size of an electron avalanche more precisely than in the previous experiment, a MWPDD with anode wire spacing b = 125μm and dimension across the wires 1cm (to diminish the parallax) was prepared. 14 hours after preparation, the results Ss =$\eta_\alpha$(c)/$\eta_\alpha$(s+c) ≈ 0.06 and maximum $\eta_\alpha$(s+c) ≈ 0.9 were obtained. Thus the transverse size of electron avalanche in this MWPDD is much smaller than 125μm. This result is very important since it means that it is possible to expect that detectors with smaller pitch may have better coordinate resolutions than 125μm.

**Experiment # 4.** *Study of MWPDD with a large pitch*

Parameters of the experiment                                                                    Table 5

| Detector | A, cm$^2$ | b, μm | d, cm | L, mm | I, cm$^{-2}$·s$^{-1}$ | t,$^0$C | Impurity of CsI |
|---|---|---|---|---|---|---|---|
| MWPDD | 2.2×2.2 | 750 | 12 | 0.8 | 9.1 | 18 | 10$^{-4}$ |

To develop a theory of the EDM in PD under the action of the external electric field it is very important to know, what spatial sensitivity the detector has right after assembly. For this, it is necessary to use a MWPDD with a b of the same magnitude as the avalanche size. In this case investigations of the spatial resolution of the MWPDD were carried out with pitch b = 750 μm. As in the first experiment, measurements have started in one hour after assembling of the detector. The investigations have shown that $\eta_\alpha$(s+c) and Ss were 100% and 0.6 respectively, i.e. in one hour after assembly porous detector had quite poor spatial resolution. As was expected, on the second day it had Ss≈0 and operated stably.

**Experiment # 5.** *Study of MWPDD kept at temperature t=31$^0$C.*

Parameters of the experiment                                                                    Table 6

| Detector | A, cm$^2$ | b, μm | d, cm | L, mm | I, cm$^{-2}$·s$^{-1}$ | T,$^0$C | Purity of CsI |
|---|---|---|---|---|---|---|---|
| MWPDD | 2.2×2.2 | 210 | 12 | 0.8 | 8.8 | 31 | 10$^{-4}$ |

For the development of the theory of the EDM in PD it is also very important to know the dependence of stabilization time on the conditions of the experiment, such as the temperature T at which the porous detector was kept after assembly. It was shown, that for tested detectors the stabilization time depends on the temperature T. Studies were carried out at different values of T, however results of only T=31$^0$C are presented. The dependences of alpha particle detection efficiencies $\eta_\alpha$ (s+c) and $\eta_\alpha$(c) on U in one hour after porous CsI layer deposition are illustrated in fig.11. As is visible from this figure, as distinct from the results presented in fig.6, which were also obtained in one hour after assembly, but at T=18$^0$C, in the case of T=31$^0$C one hour is enough for acquisition by MWPDD of good spatial sensitivity. Thus processes bringing to a stabilization of MWPDD properties are accelerating with the increase of temperature T at which the porous detector was kept after assembly.

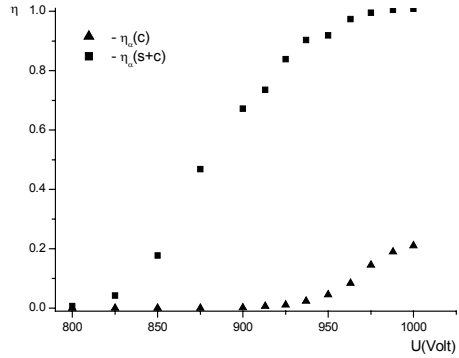

**Fig. 11** U dependences of the total detection efficiency of α – particles $\eta_\alpha(s + c) = \eta_\alpha(s) + \eta_\alpha(c)$ and the efficiency of simultaneous detection by the both groups of anode wires $\eta_\alpha(c)$ in one hour after assembly. This detector was kept after assembly at temperature $T = 31^0C$.

**Experiment # 6.** *Study of MWPDD at intensity of α - source 711cm$^{-2}$ s$^{-1}$.*

Parameters of the experiment                                      Table 7

| Detector | A, cm$^2$ | b, μm | d, cm | L, mm | I, cm$^{-2}\cdot$s$^{-1}$ | T,$^0$C | Impurity of CsI |
|---|---|---|---|---|---|---|---|
| MWPDD | 2.2×2.2 | 250 | 12 | 0.8 | 711 | 24 | 10$^{-4}$ |

All previous detectors have been studied at very low intensities of the incident particles and had CsI with purity 99.99. Since polarization effects should increase with the increase of the beam intensity, CsI with purity 99.999 was used for research of the detector at α –particles intensity $I = 711$cm$^{-2}$ s$^{-1}$. The reason was that polarization phenomena are influenced by the cleanliness of the medium. The measurements started one hour after the assembly of the detector. At the beginning again as with lower intensities, the detector had poor spatial resolution, however in few hours the spatial resolution improved, Ss fell down to Ss = 0.13 and detector starts to operate stably. The results of time stability measurements of $\eta_\alpha(s+c)$ and $\eta_\alpha(c)$ are presented in fig.12.

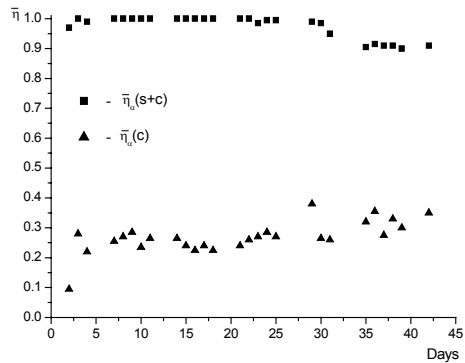

**Fig. 12** The time stability of total detection efficiency of α – particles $\eta_\alpha(s + c) = \eta_\alpha(s) + \eta_\alpha(c)$ and the efficiency of simultaneous detection by the both groups of anode wires $\eta_\alpha(c)$ of MWPDD obtained for rate $I = 711$cm$^{-2}$ s$^{-1}$. The detector after assembly was kept in vacuum at temperature $24^0$C.

Note that MWPDD was switched off at nights and during weekends. One can clearly see that with high accuracy both $\eta_\alpha(s+c)$ and $\eta_\alpha(c)$ have constant values during 30 days, and $\eta_\alpha(c)/\eta_\alpha \approx 0.3$, but on the 31-st day the $\eta_\alpha$ decreases slightly and then again remains constant. A higher value of Ss = $\eta_\alpha(c)/\eta_\alpha$ compared with those of other detectors is not a consequences of the transverse size of electron shower in porous layer, but is due to a small distance from detector to alpha particle source and large area of α – source (≈ 0.8cm), and accordingly a larger parallax. Note, that the reason for the decrease of $\eta_\alpha$(s+c) after thirtieth day is not clear to us.

Thus, one can conclude in two days after assembling MWPDD operates at intensity 711cm$^{-2}$ s$^{-1}$ of heavily ionizing α –particles as good as at lower rates.

**Experiment # 7.** *Study of larger MWPDD.*

Parameters of the experiment                                                         Table 8

| Detector | A, cm$^2$ | b, μm | d, cm | L, mm | I, cm$^{-2}$·s$^{-1}$ | T, $^0$C | Impurity of CsI |
|----------|-----------|-------|-------|-------|------------------------|----------|-----------------|
| MWPDD    | 5×5       | 250   | 16    | 0.7   | 5.1                    | 18       | 10$^{-4}$       |

All above presented results were obtained for MWPDDs having a small effective area. Such detectors are not suitable for using in real experiments. Those investigations aimed to establish the peculiarities of the porous detectors. An area of 25 cm$^2$ is also not large enough, but it is suitable as a building block of a large area detectors system. Reassuring is that although the working area of studied detector is 4 times larger that in previous cases, its spatial resolution is as good and it performance is as stable as those of small area detectors. Figure 13 shows the time stability of the alpha particle detection efficiencies $\eta_\alpha$(s+c) and and $\eta_\alpha$(c) over a period of 3-42 days.

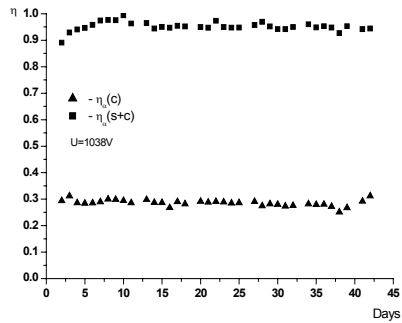

**Fig. 13** The time stability of total detection efficiency of α - particles $\eta_\alpha$(s + c) = $\eta_\alpha$(s) + $\eta_\alpha$(c) and the efficiency of simultaneous detection by the both groups of anode wires $\eta_\alpha$(c) of MWPDD with area 5x5 cm$^2$ obtained over a period of 3-42 days operation.

On the third and forth days this detector operated continuously, while on remaining days it was switched off at nights and during weekends. One can clearly see that $\eta_\alpha$(s+c) increased slowly during first six days and then had a constant values of $\eta_\alpha$ = 0.95 and $\eta_\alpha$(c) = 0.3 (Ss = 0.32).

The higher value of Ss for this detector as compared with those of previous detectors can be explained by the fact that effective area of this detector is larger than in former detectors and consequently the parallax is larger (about 40μm).

**Experiment # 8.** Registration of X-rays.

Parameters of the experiment                                                         Table 9

| Detector | A, cm$^2$ | b, μm | d, cm | L, mm | I, cm$^{-2}$·s$^{-1}$ | T, $^0$C | Imurity of CsI |
|----------|-----------|-------|-------|-------|------------------------|----------|----------------|
| MWPDD    | 2.2×2.2   | 250   | 15    | 1.1   | 12.3                   | 18       | 10$^{-4}$      |

The energy released by X- quanta in a porous CsI layer is compatible with energy losses of minimum ionizing particles in this layer, thus these results are indicative of using a MWPDD for the registration of minimum ionizing particles.. As in the case of α-particles, X-rays at the beginning had poor spatial resolution and bad stability, but day by day, the counting characteristics improved. 16 hours after the preparation of the MWPDD, X-ray detection efficiency $\eta_x(s+c)$ rose and then plateaued at 47 %, while $\eta_x(c)$ was very small (Ss << 1) and did not grow noticeably (fig.14). At that time, the spatial resolution of the detector was not worse than the anode-wire spacing b = 250 μm.

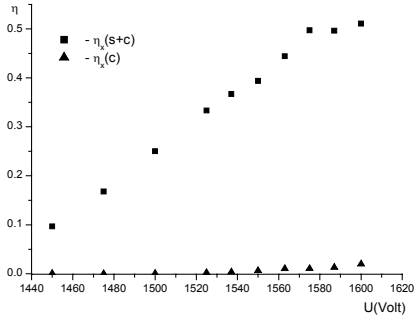

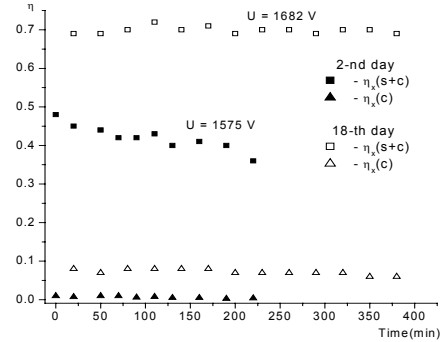

**Fig. 14**

**Fig. 15**

**Fig. 14** U dependences of the total detection efficiency of X- rays $\eta_x(s + c) = \eta_x(s) + \eta_x(c)$ and the efficiency of simultaneous detection by the both groups of anode wires $\eta_x(c)$ obtained in 16 hour after porous detectors assembly. The detector after assembly was kept in vacuum at temperature $18^0$C.

**Fig. 15** The time stability of total detection efficiency of X-rays $\eta_x(s + c) = \eta_x(s) + \eta_x(c)$ and the efficiency of simultaneous detection by the both groups of anode wires $\eta_x(c)$ of MWPDD obtained on the 2-nd and 18-th days operation.

The time-stability measurements of this detector started on the second day after the MWPDD assembly and lasted for 29 days. During the first 18 days the detector was detecting X-rays, then from the $19^{th}$ to $28^{th}$ days it detected α-particles, and on the last ($29^{th}$ day) it again detected X-rays. In all cases its performance was good. The results obtained for detector's operation for 3.6 hours on the second day at U = 1475V, for 12 hours on 16-th day at U= 1550V and for 12 hours on 18-th day at U=1682V are demonstrated in fig.15. One can see, that on the second and 16-th days at the beginning $\eta_x(s+c) \approx 0.5$, however on the second day $\eta_x(s+c)$ drops in the course of time, while on $16^{th}$ day it is practically constant. However on the 18-th day, when the voltage was higher, the $\eta_x(s+c)$ increases to $\approx 0.7$ and then within the experimental errors doesn't change. Given that in the detector used, the CsI layer absorbs only ≈70% of incident X-rays, the MWPDD's detection efficiency of absorbed X- rays can in principle reach 96%.

6.**2 Investigation of microstrip porous dielectric detector**

6.2.1. Experiment # **9. Registration of α-particles.**

Parameters of the experiment                                      Table 10

| Detector | A, cm$^2$ | b, μm | d, cm | L, mm | I, cm$^{-2}\cdot$s$^{-1}$ | T,$^0$C | Impurity of CsI |
|---|---|---|---|---|---|---|---|
| MSPDD | 2.2×2.2 | 100 | 16 | 0.8 | 5.1 | 20 | $10^{-4}$ |

The tested MSPDD was switched on the fourth day after assembly. Figure 16 presents the variation with the MSPDD anode voltage U of the detection efficiencies $\eta_\alpha(s+c)$ and $\eta_\alpha(c)$ obtained on the fourth and thirty second days of operation. We see that on both days $\eta_\alpha(s+c)$ rises very sharply and then plateaus. The voltage corresponding to a plateau on the thirty second day is ≈100 V higher than that on the fourth day, but in the region of plateau detection efficiencies are the same and for both days Ss ≈ 0.1. Thus the spatial resolution of the MSPDD operation on the forth and on the thirty second day after assembly the width of the electron shower in the porous layer of MSPDD is much smaller than the pitch b = 100 μm and its spatial resolution is equal to 100 μm. The results on the time stability measurements on the 4-th and on the 32-nd days are given in fig.17. We see, that the stability on the 32-nd day is significantly better than on the fourth day, i.e. the stability of the detector improves during operation.

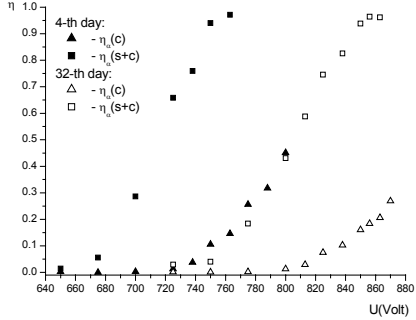 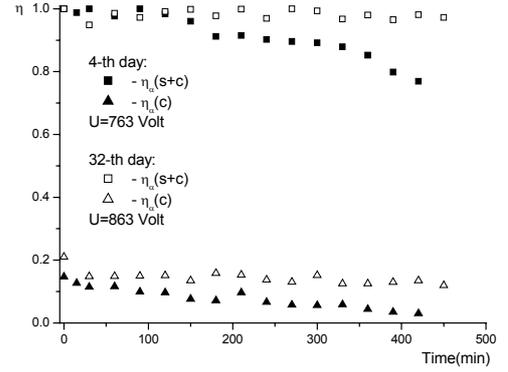

**Fig. 16**  **Fig. 17**

**Fig. 16** U dependences of the total detection efficiency of α – particles $\eta_\alpha(s + c) = \eta_\alpha(s) + \eta_\alpha(c)$ and the efficiency of simultaneous detection by the both groups of anode wires $\eta_\alpha(c)$ for MSPDD obtained in 4-th and 32-nd days operation. The detector after assembly was kept in vacuum at temperature $20^0C$ and in first three days MSPDD did not switched on.

**Fig. 17** The time stability of total detection efficiency of α - particles $\eta_\alpha(s + c) = \eta_\alpha(s) + \eta_\alpha(c)$ and the efficiency of simultaneous detection by the both groups of anode wires $\eta_\alpha(c)$ of MSPDD obtained on the 4-th and 32-nd days operation. Investigations are made just after finishing the measurements results of which are demonstrated in figure 16.

Unfortunately, we could not detect X-rays by this MSPDD, because the breakdowns between the grid cathode and strips did not allow applying a higher voltage necessary for the detection of X-rays. The aluminized Mylar cathode was also tested. Although it provided high homogeneity of electric field near the cathode, but our investigations showed that the MSPDD with aluminized Mylar cathode dose not provide a good X- ray detection efficiencies in DC mode operation and the perform unstable.

**Discussion**.

It was found experimentally that porous multiwire and microstrip detectors immediately after the thermal deposition of the CsI layers 1. operate unstably and have very poor spatial sensitivity, 2. after being kept in vacuum for a certain time acquire high stability and good spatial resolution, 3. after acquisition of these properties characteristics of detectors do not change. It is possible to explain the first and second effects as follows: just after preparation a porous CsI layer crystallization processes in the pore walls are not completed and high densities of bulk and surface defects exist in them. These defects are charge-carrier trapping centers and form high space and surface charges, i.e. dielectric media is polarized. The polarization fields changes make the resulting electric field weaker and not directed to the wires (strips). These effects suppress the process of electron drift and multiplication, widen the electron shower and thus deteriorate the spatial resolution of the detector. The second effect can be explained by assuming that the crystalline structure of the pore walls spontaneously improves, which results in a decrease of the densities of defects and accordingly to a fall of the degree of dielectric media polarization. The third effect is clearer, since after completion of crystallization processes the crystalline structure does not change. This assumption is in agreement with the fact, that the time of the improvement of characteristic of the detector doesn't depend on whether it is on or off, and depends only on the temperature of the chamber in which porous detector is placed after assembling.

Note, that in used detectors α particle produces $7 \cdot 10^4$ electron-hole pairs, while X-quanta produce only $3 \cdot 10^2$ [62]. Since the registration threshold was the same in both cases, this means that in our experiments the detectors gain for heavily ionising particles is much less than that for weakly ionising particles.

Unfortunately we didn't have α-particle sources with rate higher than I = 711cm$^{-2}$s$^{-1}$ and understand that this rate is significantly lower than fluxes of particles in modern experiments. It is possible that for normal operation of porous detectors at higher intensities of incident particles if will be necessary to use cleaner active materials than those used in this work (semiconductor detectors have a purity up to of ~ $10^{-10}$). The fact that MWPDD has good coordinate resolution, operates stably for 5 MeV α –particles at a flux of 700 cm$^{-2}$ s$^{-1}$ and registers 5.9 keV X-rays with sufficiently high effectiveness leads us to believe that such detectors can also be used to detect minimum ionizing particles with higher fluxes.

Note that after completing all of the measurements, the detectors were disassembled, and the surfaces of porous CsI were investigated with an optical microscope. No changes in the state of the CsI porous layer were observed i.e. the detectors could be still operable.

**Conclusions**.

We now can conclude that multiwire and microstrip porous dielectric detectors operated in the DC mode perform stably and the width of the electron shower in these detectors is much smaller than 125μm and 100 μm respectively. These results mean, that detectors with smaller pitch may have coordinate resolutions better than 100μm. Some stability and spatial resolution problems due to polarization of the porous material are encountered from the very beginning, however they disappears in due time after assembly of porous detectors. The time of such changes of porous detector properties depends on the temperature at which it is held after assembly. Then, after stabilization of porous detectors, their properties do not change independent on whether they perform or not.

It is necessary to note that the available experimental data are insufficient for a final quantitative interpretation of the discovered effect and it is necessary to perform more thorough experimental and theoretical studies. We hope that after publishing this article the interest to that problem will increase and others groups will also start to study these detectors.

*Acknowledgements*

The author is grateful to the founders of the International Science and Technology Center, to collaborators of ISTC proposal project Professors M. A. Piestrup and Ch.K.Gary for valuable collaboration, to ISTC officers L. Horowicz and S. Temeev for their support. The author also expresses his gratitude to G. Ayvazyan, R. Aivazyan, H. Vardanyan, for their assistance in the process of this study and professors F. Sauli and A.K.Odian for rendering of the large support, G. Grigoryan for help in data processing and G. Tadevosyan for their help in the preparation of this paper.